\documentclass[aps,reprint,longbibliography,superscriptaddress]{revtex4-1}
\usepackage{amsmath}
\usepackage{amssymb}
\usepackage{bm}
\usepackage{epsfig}
\usepackage{graphicx}
\usepackage[dvipsnames]{xcolor}
\usepackage{color}
\usepackage[unicode=true,colorlinks=true,citecolor=blue,urlcolor=blue]{hyperref}
\usepackage[english]{babel}
\usepackage{braket}

\renewcommand{\Re}{\mathop{\rm Re}}

\newcommand{\e}{\mathrm{e}}

\renewcommand{\i}{{\rm i}}
\renewcommand{\d}{\mathrm d}
\renewcommand{\emph}{\textit}

\renewcommand{\braket}[1]{\left\langle #1 \right\rangle}

\begin{document}

\title{Electric current noise in mesoscopic organic semiconductors}

\author{D.~S.~Smirnov}
\author{A.~V.~Shumilin}
\email{hegny@list.ru}
\affiliation{Ioffe Institute, 194021 St. Petersburg, Russia}

\date{\today}

\begin{abstract}
We demonstrate that nuclear spin fluctuations lead to the electric current noise in the mesoscopic samples of organic semiconductors showing the pronounced magnetoresistance in weak fields. For the bipolaron and electron-hole mechanisms of organic magnetoresistance, the current noise spectrum consists of the high frequency peak related to the nuclear spin precession in the Knight field of the charge carriers and the low frequency peak related to the nuclear spin relaxation. The shape of the spectrum depends on the external magnetic and radiofrequency fields, which allows one to prove the role of nuclei in magnetoresistance experimentally.
\end{abstract}

\maketitle

\section{Introduction}
\label{sec:intro}

Organic semiconductors represent relatively new class of semiconductors, which attract increasing interest nowadays. Although they are already successfully used in light emitting diodes~\cite{OLED,SmallMolOLED}, organic solar cells~\cite{SolCells,Zhou2019,SmallMolOSC2020} and other devices, their transport properties are not completely understood theoretically yet. Organic semiconductors are amorphous materials, which consist of single molecules or short polymers. The transport in them typically operates via hopping of polarons between molecular orbitals \cite{Baessler2012,Bassler}. It is quite similar to the hopping conductivity in inorganic semiconductors~\cite{Shk}. For this reason in this paper we use the notations of electrons, holes and hopping sites.

A unique feature of organic semiconductors is the strong coupling between electric current and nuclear spins. It was shown experimentally back in 2003 that light emission from organic diodes can be significantly modified by application of magnetic fields as small as $100$~mT~\cite{kalin}. Then in 2005 it was found that the resistivity of organic semiconductors can be affected by the magnetic fields in the same range~\cite{OMAR0}. This effect is called organic magnetoresistance (OMAR). It takes place in a number of different organic materials at liquid helium as well as at room temperatures.

Qualitatively, OMAR is related to the suppression of the electron and hole spin relaxation caused by the hyperfine interaction with atomic nuclei~\cite{prigodin,bobbert}. Organic semiconductors are nonmagnetic materials, so at small magnetic fields the average electron and nuclear spin polarizations are negligible. However, there are unavoidable nuclear spin fluctuations, which create stochastic Overhauser field for the electrons. Due to this,  even in zero external magnetic field, the electron spin precesses with random frequency between the hops, which results in the spin relaxation~\cite{schulten,merkulov02}. By contrast, in the strong magnetic field, the electron spin precession frequency equals to Larmor frequency which is the same for all hopping sites, so the spin relaxation gets suppressed~\cite{PRC}. Recently, some of us have shown that OMAR can be related to the nonequilibrium electron spin correlations~\cite{AVS2018,AVS2020}, which appear due to the applied voltage. The relaxation of these correlations leads to OMAR.


Nevertheless, up to date there is no unambiguous experimental proof of the nuclear origin of OMAR. In the same time, the spin-orbit interaction can play an important role in the hopping conductivity regime~\cite{KKavokin-review,Hopping_spin,PhysRevB.98.155304}. In Refs.~\cite{SO-OMAR1,SO-OMAR2} it was suggested as an origin of OMAR. Some other alternatives have been also suggested~\cite{kabanov2012}. Therefore, it is desirable to propose an experiment, which can evidence the role of nuclear spins. In this paper we suggest to measure the current noise spectra in mesoscopic organic semiconductors. Although the nuclear spins are often assumed to be static~\cite{bobbert,HF2012-1,HF2012,larabi,AVS2020}, their dynamics unavoidably takes place due to the interaction with the electrons. This dynamics is slow and does not change the average electric current. However it leads to the fluctuations of the current in the mesoscopic samples at the frequencies determined by the nuclear spin dynamics. These fluctuations can be suppressed by the external magnetic field similarly to OMAR, which allows one to separate them from shot and $1/f$ noises.  In this paper we calculate the current noise spectrum and demonstrate, that its measurement will allow one to prove experimentally the importance of the hyperfine interaction in OMAR.

The paper is organized as follows. In the next section we describe the two alternative microscopic mechanisms of OMAR and establish for them a common relation between current and nuclear spin correlations. Then in Sec.~\ref{sec:i_noise} we calculate the current noise spectra in mesoscopic organic semiconductors. In Sec.~\ref{sec:concl} we discuss the limits of applicability of our theory and summarize our findings.

\section{Relation between nuclear spin dynamics and resistivity}
\label{sec-NS-cur}


OMAR is caused by the dependence of the resistivity on the spin relaxation time. Microscopically, there are two main mechanisms of this dependence: (i)  In the \textit{bipolaron mechanism}, it is assumed that a single hopping site can be occupied by two electrons or two holes only if they are in the singlet state due to the strong exchange interaction~\cite{bobbert}. (ii) The \textit{electron-hole mechanism} is based on the spin dependent recombination of electrons and holes~\cite{prigodin}. Its rate is assumed to be different for the singlet and triplet states of electron hole pair. In both mechanisms, it is necessary to take into account correlations between spins of the charge carriers to describe OMAR~\cite{AVS2015,AVS2018,AVS2020}. Below we present the common model for the description of the conductivity in mesoscopic sample and then calculate the electric current for the electron-hole and bipolaron mechanisms of OMAR. We focus our discussion on the mesoscopic samples because the current noise in them is the strongest.

The distribution of the hopping rates in organic semiconductors is exponentially broad due to the following reasons: (i) The overlap integrals between neighboring hopping sites differ by several orders of magnitude \cite{Masse2017}. (ii) the typical width of distribution of site energies in organic semiconductors is $0.1$~eV, which is much larger than the thermal energy at room temperature. Therefore, the transport in organic semiconductor can be described by the precolation theory \cite{Shk}. It predicts that the so-called precolation cluster would carry most of the current. The cluster consists of pairs of sites with hopping rates faster than or comparable with the critical hopping rate. Most of the hops in the percolation cluster are much faster than the critical rate and are not essential for the calculation of conductivity. The conductivity is controlled by the rare ``critical'' pairs of sites in the percolation cluster where the hopping rates are comparable to the critical rate. The typical distance between these pairs is called a correlation length of the percolation cluster $L_c$. If the size of a sample of organic semiconductor $L$ is smaller than $L_c$, its conductivity is controlled by a single critical pair of sites. In this case, the current noise in the sample is the strongest. Note that mesoscopic sample can still contain large number of hopping sites because $L_c$ is much larger than the typical distance of a single hop \cite{Shk}.

\begin{figure}[t]
    \centering
        \includegraphics[width=0.5\textwidth]{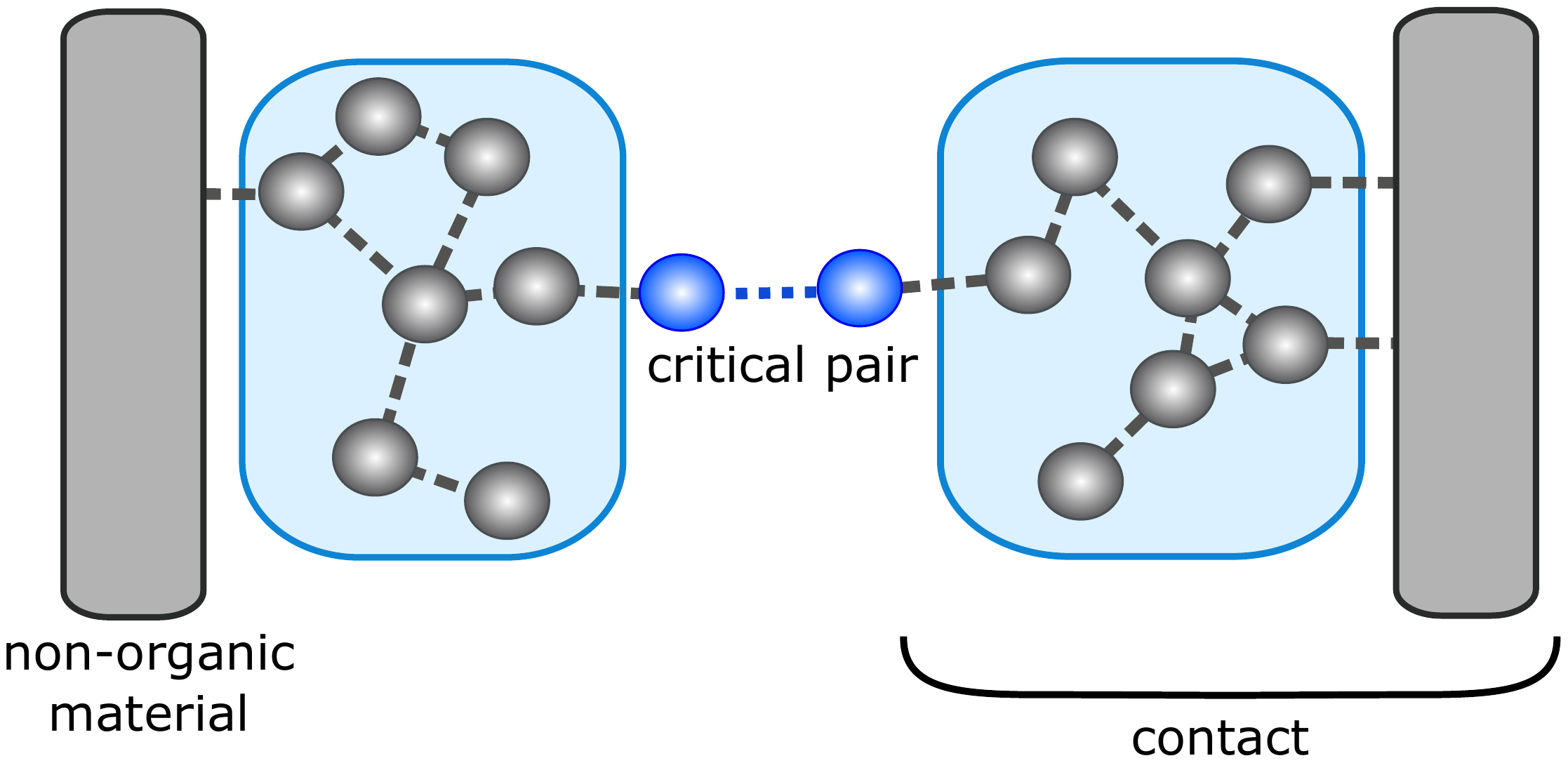}
        \caption{The structure of a mesoscopic sample. Inorganic contacts (grey areas) are connected to the parts of the percolation cluster with relatively high conductivity. Together they represent an effective contact for the critical pair of sites that controls the conductivity of the organic layer.}
    \label{fig:perc}
\end{figure}

The structure of a mesoscopic sample is shown in Fig.~\ref{fig:perc}. The inorganic contacts are connected to the parts of the percolation cluster with relatively high conductivity. These parts of percolation cluster can be considered as parts of the contacts for the critical pair of sites that controls the conductivity of the sample. The local chemical potentials are formed in these parts of the percolation cluster. In reality they can differ from chemical potentials in inorganic contacts due to the carrier injection. This situation as well as the underlying physics is similar to the double quantum dot system~\cite{hanson07,PhysRevB.100.075409}.

To describe the effect of the nuclear spins on the current, the correlations of electron and hole spin directions should be taken into account. For many charge carriers, there are extremely many correlations. The effect of different spin correlations was studied in details in Ref.~\cite{AVS2020} for the bipolaron mechanism of OMAR in close to equilibrium conditions. It was found that in the materials where the shape of OMAR is close to Lorentzian~\cite{OMAR0}, it is enough to take into account the correlations between the spins in the closest pairs of sites only. We adopt this approximation and consider only the spin correlations in the critical pair of sites. We assume that the occupation numbers and spin directions of the sites in the organic parts of the contacts are not correlated with occupation numbers and spins in the critical pair and between themselves. The averaged product of filling numbers is equal to the product of averaged filling numbers when the correlations are neglected. The averaged products of spin components are equal to zero without the correlations, because we consider organic semiconductors at high enough temperature.

\begin{figure}[t]
  \centering
  \includegraphics[width=0.9\linewidth]{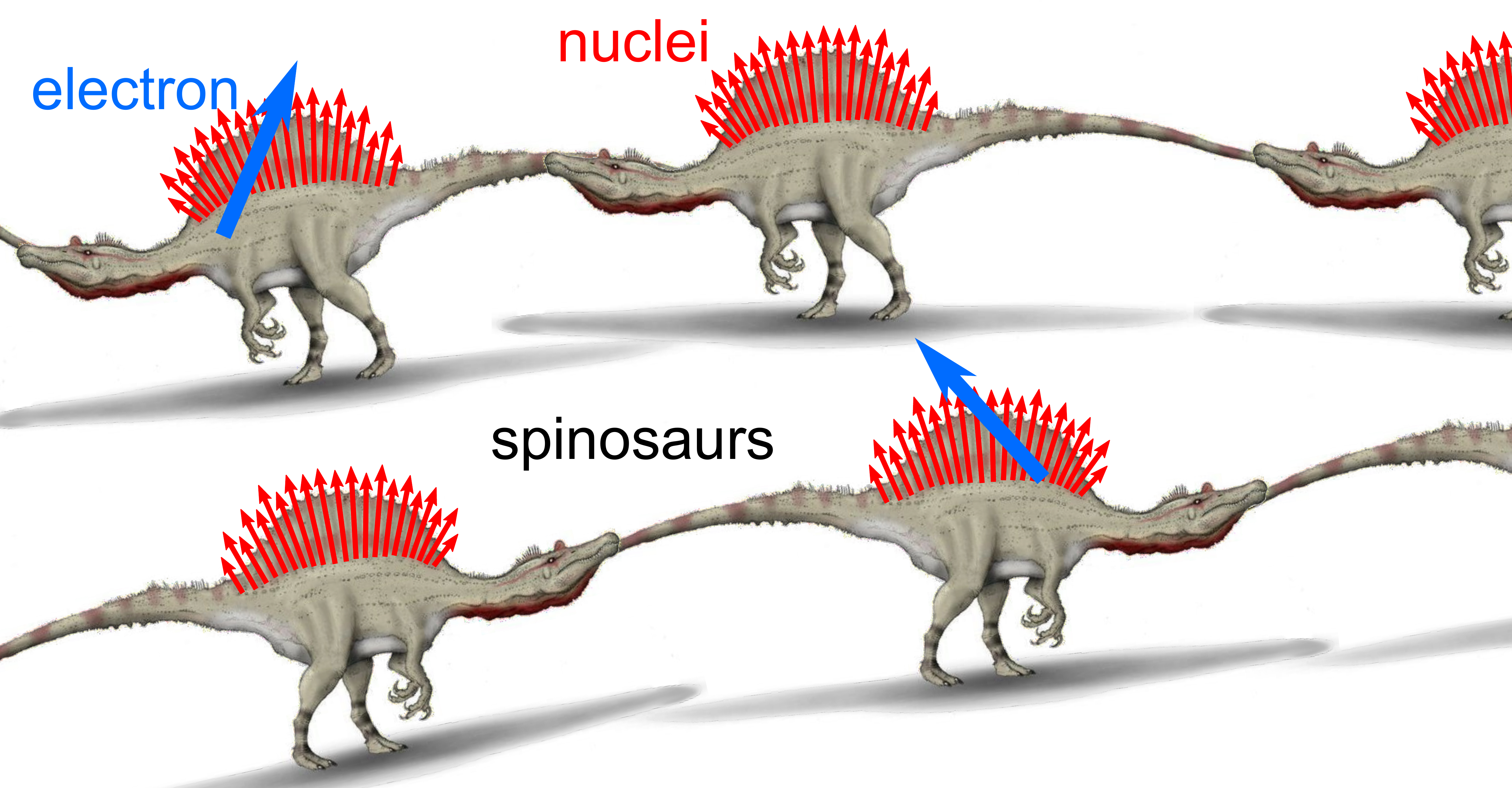}
  \caption{Impression of the electron spin (blue arrow) hopping between complex organic molecules (spinosaurs), which carry nuclear spins (red arrows). The orientation of nuclear spins in the equilibrium is random.}
  \label{fig:spins}
\end{figure}

The spins of electrons and holes in the critical pair of sites interact with atomic nuclei. Typical hopping sites (molecular orbitals) contain dozens of nuclear spins. For example, the molecule of Alq$_3$ contains eighteen hydrogen atoms with the nuclear spins $1/2$, three nitrogen atoms with the spins $1$, and one aluminum atom with the spin $5/2$. As a result, a single electron spin can interact with many nuclear spins at each hopping site. The distribution of the coupling constants strongly depends on the electron wave function~\cite{book_Glazov}. In this work, we abstain from the description of the complex structure of the hopping sites, and imagine them as some ``spinosaurs'' carrying nuclear spins, see Fig.~\ref{fig:spins}. The electrons and holes hop over the backs of the spinosaurs and uniformly interact with the nuclear spins, which is usually called a box model. For this oversimplified model the compact expressions for the nuclear spin dynamics were derived in Ref.~\cite{my_box}. They will be used below to describe the electric current fluctuations.

We assume that the charge carrier spins at the different sites of the critical pair interact with the different nuclei. At each site the electron spin precession frequency $\bm\Omega_e$ is composed of the spin precession frequency in the external magnetic field $\bm\Omega_B$ and the precession frequency in the fluctuation of the nuclear field $\bm\Omega_N$:
\begin{equation}
  \label{eq:Omega_e}
  \bm\Omega_e=\bm\Omega_B+\bm\Omega_N.
\end{equation}
The nuclear spin precession frequency slowly fluctuates with time and at the time scale of the electron dynamics can be considered as frozen. It is described by the probability distribution function
\begin{equation}
  \label{eq:F}
  \mathcal F(\bm\Omega_N)=\frac{1}{(\sqrt{\pi}\delta)^3}\exp\left(-\Omega_N^2/\delta^2\right),
\end{equation}
where parameter $\delta$ describes the dispersion.

An important feature of organic semiconductors is the vanishing concentration of the resident charge carriers in the equilibrium. The electrons are injected from the contacts, so the usual linear response theory is typically unacceptable for organic semiconductors~\cite{Baldo}. For this reason we consider the nonlinear regime of the conductivity. We will show below that for the both mechanisms of OMAR, the current has the form
\begin{equation}\label{cur-form}
  J=J_0+J_1 \cos^2(\theta_{12}).
\end{equation}
Here $J_0$ is the contribution independent of nuclear spins, $\theta_{12}$ is the angle between the spin precession frequencies $\bm\Omega_e$ at the critical pair of sites, and $J_1$ describes the contribution sensitive to the magnetic field. The microscopic expressions for $J_0$ and $J_1$ will be obtained below for the nonlinear conductivity regime. The fluctuations of $\theta_{12}$ lead to the electric current noise.


\begin{figure}[t]
    \centering
        \includegraphics[width=0.5\textwidth]{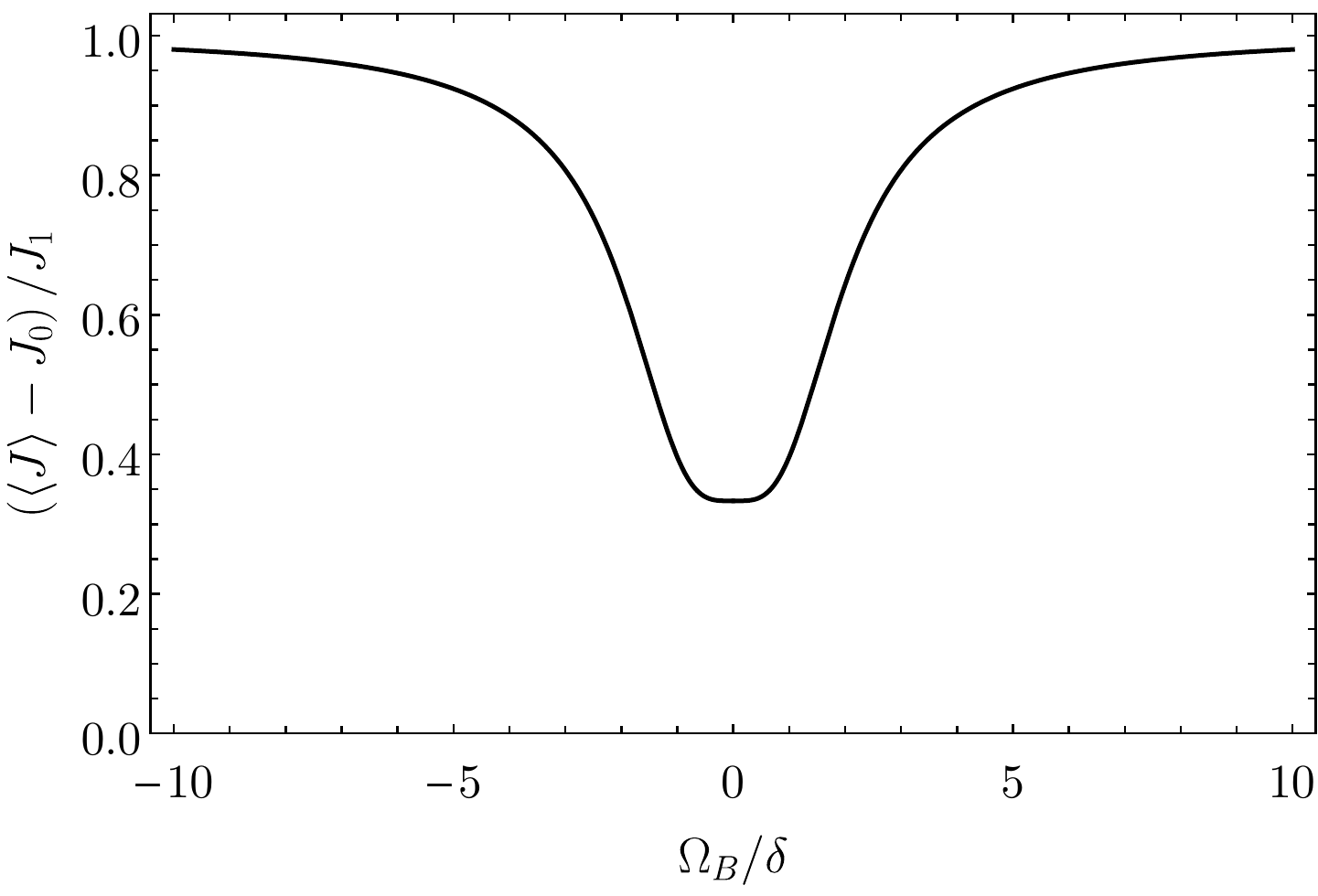}
        \caption{The magnetic field dependent contribution to the electric current [$\left\langle\cos^2(\theta_{12})\right\rangle$] as a function of the Larmor precession frequency calculated after Eq.~\eqref{eq:cos2}.}
    \label{fig:Dawson}
\end{figure}

Using the distribution function~\eqref{eq:F} we find the average electric current
  \begin{equation}
    \left\langle J \right\rangle = J_0 + J_1 \left\langle \cos^2(\theta_{12}) \right\rangle,
  \end{equation}
  where
  \begin{multline}
    \label{eq:cos2}
    \left<\cos^2(\theta_{12})\right> = \frac{1}{3} \\ + \frac{3}{2}\left(\frac{\delta}{\Omega_B}\right)^6\left[\frac{2}{3}\left(\frac{\Omega_B}{\delta}\right)^3-\frac{\Omega_B}{\delta}+3D\left(\frac{\Omega_B}{\delta}\right)\right]^2
  \end{multline}
  with $D(x)=\exp(-x^2)\int_0^x\exp(y^2)\d y$ being the Dawson integral. This expression is shown in Fig.~\ref{fig:Dawson}. One can see that the external magnetic field of the order of $\delta$ changes the electric current by the value of about $J_1$. Typically in the experiments $J_1/J_0\sim0.1$~\cite{OMAR0}. 


\subsection{Bipolaron mechanism}
\label{sec:bip_mech}

\begin{figure}[tbp]
    \centering
        \includegraphics[width=0.5\textwidth]{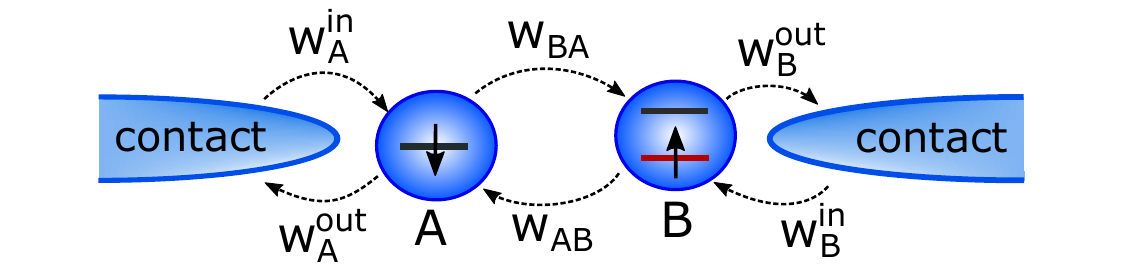}
        \caption{The bipolaron mechanism of OMAR. The $B$ site is always occupied with an electron, which is indicated by the red energy level. The arrows and labels show the possible hops and the corresponding rates.}
    \label{fig:mechAB}
\end{figure}

The bipolaron mechanism is related to the possibility of double occupation of a single hopping site by two electrons or two holes in the singlet state only. To be specific we will consider the electrons. Since the electron spin is conserved during the hop, the hopping from a singly occupied site to another singly occupied site is possible only when the spins are in the singlet state. The detailed theory of bipolaron mechanism of OMAR including all the possible spin correlations in systems close to equilibrium was developed in Ref.~\cite{AVS2020}. This theory involves two types of hopping sites: A-type sites that are never doubly occupied and B-type sites that can be doubly occupied but never lose the last electron. In this work we adopt this model to consider a mesoscopic sample where the critical pair consists of $A$ and $B$ sites (Fig.~\ref{fig:mechAB}). In contrast to the previous model we do not assume the system to be close to the equilibrium.

The dynamics of the spin correlations in the critical pair is described by the following equation~\cite{AVS2020}:
\begin{equation}
  \label{srel_bp}
  \frac{d \overline{s_A^\alpha s_B^\beta}}{dt} = -R_{\alpha\beta;\alpha'\beta'} \overline{s_A^{\alpha'} s_B^{\beta'}} - \frac{J_{AB}}{4e} \delta_{\alpha\beta}
\end{equation}
Here the sum over the repeating indices is assumed, $\overline{s_A^\alpha s_B^\beta}$ is the quantum mechanical average of the product of components of spins at sites $A$ and $B$,
\begin{equation}
  J_{AB} = 2 e W_{AB}\overline{n}_B - e W_{BA}\frac{\overline{n}_A - 4\overline{s_A^\alpha s_B^\alpha}}{2}
\end{equation}
is the current between $A$ and $B$ sites with $\overline{n}_A$ and $\overline{n}_B$ being the average probabilities of single and double occupation of these sites and $e$ being the electron charge. The notations of the hopping rates are introduced in Fig.~\ref{fig:mechAB}. Due to the small concentration of electrons in organic semiconductors we take $\overline{n_A n_B} = 0$.

The last term in Eq.~\eqref{srel_bp} describes the generation of the spin correlation due to the spin dependent hopping. Another term describes the dynamics and relaxation of the spin correlations due to the hyperfine interaction and hopping:
\begin{multline}\label{R-bp}
R_{\alpha\beta;\alpha'\beta'} =  W_A^{out} \delta_{\alpha\alpha'}\delta_{\beta\beta'} \\ + \frac{W_{BA}}{2} (\delta_{\alpha\alpha'}\delta_{\beta\beta'} - \delta_{\alpha\beta'}\delta_{\beta\alpha'}) \\
-  \left( \epsilon_{\alpha\gamma\alpha'}\delta_{\beta\beta'} \Omega_{e,\gamma}^{(A)} + \epsilon_{\beta\gamma\beta'}
\delta_{\alpha\alpha'} \Omega_{e,\gamma}^{(B)} \right)
\end{multline}
Here $\bm\Omega_{e}^{(A)}$ and $\bm\Omega_{e}^{(B)}$ are the spin precession frequencies at the corresponding sites given by Eq.~\eqref{eq:Omega_e} and $\epsilon_{\alpha\beta\gamma}$ is the Levi-Civita symbol.

In the steady state, Eq.~(\ref{srel_bp}) yields the current
\begin{equation}
J_{AB} = \frac{2 \overline{n}_B W_{AB} - \overline{n}_A W_{BA}/2}{1 + W_{BA} {\cal T}_s^{(AB)} /2},
\end{equation}
where
\begin{equation}
  \label{eq:Ts_R}
  {\cal T}_s = \delta_{\alpha\beta}\delta_{\alpha'\beta'}(R^{-1})_{\alpha\beta;\alpha'\beta'}
\end{equation}
is the effective relaxation time of the spin correlations. In the steady state, the current can be also calculated from the kinetic equations
\begin{subequations}
  \begin{equation}
    J_{AB} = e\overline{n}_A W_{A}^{out} -e W_A^{in} (1 - \overline{n}_A),
  \end{equation}
  \begin{equation}
    J_{AB} = e W_B^{in} (1-\overline{n}_B) - e \overline{n}_B W_B^{out}.
  \end{equation}
\end{subequations}
From these relations we find the occupancies $\overline{n}_A$, $\overline{n}_B$ and calculate the current for the given orientation of nuclear spins imprinted in ${\cal T}_s^{(AB)}$.

By the definition of the critical pair, $W_A^{out} \gg W_{BA}$. Moreover, to simplify the analysis, we consider the limit $\Omega_e^{(A,B)} \gg W_{A}^{out}$. In this case we obtain
\begin{equation}\label{bp-Ts}
{\cal T}_s = \frac{1}{W_{A}^{out}} \cos^2 \left( \theta_{AB} \right),
\end{equation}
where $\theta_{AB}$ is the angle between $\bm{\Omega}_e^{(A)}$ and $\bm{\Omega}_e^{(B)}$. 
In this limit, the contributions to the total current in Eq.~\eqref{cur-form} are
\begin{subequations}
  \label{bp-JAB}
\begin{equation}\label{bp-JAB0}
 J_0 =  e\frac{4 W_B^{in} W_{AB}W_A^{out} - W_A^{in}W_{BA}W_B^{out}}{4W_{AB}W_A^{out} +
 2W_{A}^{out}W_B^{out} + W_{BA}W_B^{out}
  },
\end{equation}
\begin{equation}\label{bp-JAB1}
J_1 = e\frac{W_{BA} W_B^{out} (W_A^{in} W_{BA}W_B^{out} - 4W_B^{in}W_{AB}W_{A}^{out})}{\left[
 4W_{AB}W_A^{out} + (2W_A^{out} + W_{BA})W_B^{out}\right]^2},
\end{equation}
\end{subequations}
and $\theta_{12} = \theta_{AB}$.

\subsection{Electron-hole mechanism}
\label{sec:eh_mech}

The electron-hole mechanism of OMAR involves the ambipolar transport. In this case, each molecule provides two hopping sites. Its highest occupied molecular orbital (HOMO) and lowest unoccupied molecular orbital (LUMO) represent the hopping sites for holes and electrons, respectively. Electrons and holes in organic semiconductors usually have spin $1/2$ due to the weak spin-orbit interaction. In the electron-hole mechanism the relation between current and charge carrier spin relaxation is provided by the spin-dependent recombination of electron-hole pairs.

\begin{figure}[t]
    \centering
        \includegraphics[width=0.5\textwidth]{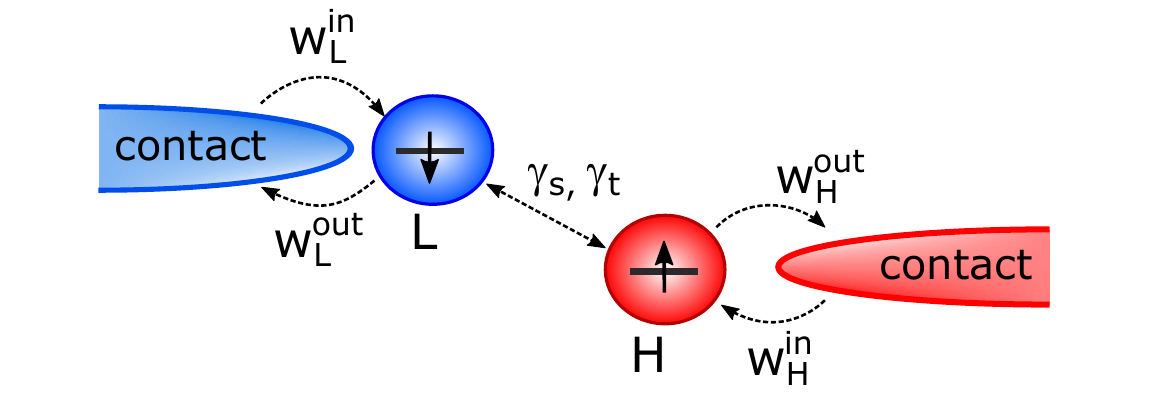}
        \caption{Illustration of the electron-hole mechanism of OMAR. LUMO sites (blue) and HOMO sites (red) can be occupied by electrons and holes, respectively. The arrows and labels show the possible hops and the corresponding rates.}
    \label{fig:mechLH}
\end{figure}

We assume that the critical pair in the mesoscopic sample is represented by LUMO site $L$ and HOMO site $H$ (Fig.~\ref{fig:mechLH}). The current in this pair $J_{LH}$ flows due to the recombination of electrons and holes. It has the form
\begin{multline}
J_{LH} = e\gamma_s \left( \frac{\overline{n_Lp_H} - 4\overline{s_L^{\alpha}s_H^{\alpha}}}{4}  \right)  \\
+ e\gamma_t
\left( \frac{3\overline{n_Lp_H} + 4\overline{s_L^{\alpha}s_H^{\alpha}}}{4}  \right),
\end{multline}
where $\gamma_s$ and $\gamma_t$ are the electron hole recombination rates for the singlet and triplet states, respectively, and the other notations are the same as in the previous subsection except for the substitution of indices $A,B$ with $L,H$. The current depends on the spin correlations, when $\gamma_s \ne \gamma_t$.

To derive the master equation for the spin correlations, we assume that the singlet and triplet recombination processes are independent. In this case we obtain
\begin{multline}\label{eh-ss}
\frac{d \overline{s_L^{\alpha}s_H^{\beta}}}{dt} = -R_{\alpha\beta;\alpha'\beta'}  \overline{s_L^{\alpha'}s_H^{\beta'}}
\\
+ (\gamma_s - \gamma_t) \frac{\overline{n_L p_H} - 4\overline{s_L^\alpha s_H^\alpha}}{16},
\end{multline}
where the relaxation matrix is given by
\begin{multline}
R_{\alpha\beta;\alpha'\beta'} = (W_L^{out} + W_H^{out} + \gamma_t) \delta_{\alpha\alpha'} \delta_{\beta\beta'} \\
+ \gamma_s ( \delta_{\alpha\alpha'} \delta_{\beta\beta'} -  \delta_{\alpha\beta'} \delta_{\beta\alpha'}) \\
-
 \epsilon_{\alpha\gamma\alpha'}\delta_{\beta\beta'} \Omega_{e,\gamma}^{(L)} - \epsilon_{\beta\gamma\beta'}
\delta_{\alpha\alpha'} \Omega^{(H)}_{e,\gamma}.
\end{multline}
Here ${\bm \Omega}_e^{(L)}$ and ${\bm \Omega}_e^{(H)}$ are the spin precession frequencies of electron at site $L$ and hole at site $H$, respectively, and the hopping rates are introduced in Fig.~\ref{fig:mechLH}.

In the steady state, from Eq.~\eqref{eh-ss} we obtain the current in the form
\begin{equation}\label{eh-cur1}
J_{LH} = e \overline{n_L p_H}\widetilde{\gamma},
\end{equation}
where
\begin{equation}
\widetilde{\gamma} = \gamma_t + \frac{\gamma_s  - \gamma_t}{4 + {\cal T}_s (\gamma_s - \gamma_t)}
\end{equation}
is the effective electron-hole recombination rate depending on the orientations of the nuclear spins through the effective relaxation time given by Eq.~\eqref{eq:Ts_R}. From kinetic equations, the current also equals to
\begin{subequations}
  \label{eh-cur2}
  \begin{equation}
    J_{LH} = e(1-\overline{n}_L)W_{L}^{in} - e\overline{n}_LW_L^{out},
  \end{equation}
  \begin{equation}
    J_{LH} = eW_H^{in}(1-\overline{p}_H) - eW_H^{out}\overline{p}_H.
  \end{equation}
\end{subequations}
To find the current these equations should be solved together with the kinetic equation for the correlation of occupancies:
\begin{equation} \label{eh-np}
(\widetilde{W}_L + \widetilde{W}_H + \widetilde{\gamma} )\overline{n_L p_H} = W_L^{in}\overline{p}_H + W_{H}^{in}\overline{n_L},
\end{equation}
where $\widetilde{W}_L = W_{L}^{in} + W_L^{out}$ and $\widetilde{W}_H = W_{H}^{in} + W_H^{out}$. This set of equations allows one to find the electric current for the given hopping rates.

When the electron spin precession is fast as compared with the hopping and recombination, $\Omega_e^{(L)}, \Omega_e^{(H)} \gg W_{L}^{out} + W_H^{out} + \gamma_t \gg \gamma_s$, we obtain
\begin{equation}\label{eh-Ts}
{\cal T}_s = \frac{\cos^2\left(\theta_{LH}\right)}{W_L^{out} + W_H^{out} + \gamma_t},
\end{equation}
where $\theta_{LH}$ is the angle between ${\bm \Omega}_e^{(L)}$ and ${\bm \Omega}_e^{(H)}$. Thus we can find the current for arbitrary $\widetilde{\gamma}$ and, therefore, for arbitrary nuclear spin directions. In this limit, the current is given by Eq.~\eqref{cur-form}, where
\begin{subequations}
\label{eh-J-12}
\begin{multline} \label{eh-J-1}
J_0 = \\ \frac{e\widetilde{\gamma}_0 W_L^{in} W_H^{in} (\widetilde{W}_L + \widetilde{W}_H)}{\widetilde{\gamma}_0(W_L^{in}\widetilde{W}_L + W_H^{in}\widetilde{W}_H)  + \widetilde{W}_L \widetilde{W}_H (\widetilde{\gamma}_0 + \widetilde{W}_L + \widetilde{W}_H) }
\end{multline}
\begin{multline} \label{eh-J-2}
J_1
= \frac{(\gamma_s - \gamma_t)^2}{16(W_L^{out} + W_{H}^{out} + \gamma_t)} \times
\\ \frac{e\widetilde{\gamma}_0 W_L^{in} W_H^{in} \widetilde{W}_L \widetilde{W}_H (\widetilde{W}_L + \widetilde{W}_H)^2}
{[\widetilde{\gamma}_0(W_L^{in}\widetilde{W}_L + W_H^{in}\widetilde{W}_H)  + \widetilde{W}_L \widetilde{W}_H (\widetilde{\gamma}_0 + \widetilde{W}_L + \widetilde{W}_H) ]^2}
\end{multline}
\end{subequations}
and $\theta_{12} = \theta_{LH}$ for the case of the electron-hole mechanism.

To summarize this section, we have calculated the electric current in the mesoscopic organic semiconductor without assumption of close to equilibrium conditions for the two mechanisms of OMAR. The current has the form of Eq.~\eqref{cur-form} and it is determined by the squared cosine of the angle between the spin precession frequencies in the critical pair of sites. In the next section we use this result to describe the electric current fluctuations.

\section{Electric current noise}
\label{sec:i_noise}

Dynamics of the nuclear spins leads to the current fluctuations in mesoscopic organic semiconductors. The absolute value of the current fluctuations in mesoscopic samples have the same order as OMAR, which can reach 10\%~\cite{OMAR0}. The current noise spectrum is given by
\begin{equation}\label{noise_g}
(\delta J^2)_\omega = \int_{-\infty}^{\infty} \left\langle  \delta J (0) \delta J (t)  \right\rangle  e^{i\omega t} dt,
\end{equation}
where $\delta J(t) = J(t) - \langle J \rangle$ is the current fluctuation and angular brackets denote the statistical averaging over the nuclear spin orientations and hops. For the bipolaron and electron-hole mechanisms, $J$ should be substituted with $J_{AB}$ and $J_{LH}$, respectively. We will study the current noise related to the nuclear spin dynamics only, and the other contributions will be briefly discussed in Sec.~\ref{sec:concl}.

For simplicity, we assume that the electron hopping is much faster than the nuclear spin precession and electron spin relaxation, which is much faster than the nuclear spin relaxation. Under these assumptions, the current noise can be described in a unified way for the bipolaron and electron-hole mechanisms. We will use notations for the bipolaron mechanism. Unless it is explicitly stated, for electron-hole mechanism the upper indices $A$ and $B$ should be substituted with the indices $L$ and $H$.

As mentioned above, for each hopping site we use the box model of the hyperfine interaction, which was solved in Ref.~\onlinecite{my_box} in the limit of many nuclear spins. The total nuclear spin dynamics, $\bm I(t)$, is described by the kinetic equation for the two component probability distribution function $f_\pm(t,\bm I)$:
\begin{equation}
  \label{eq:kinetic}
  \frac{\partial f_\pm}{\partial t}+\bm\nabla\left[\left(\bm\omega_n^\pm\times\bm I-\frac{\bm I}{\tau_s^n}\right)f_\pm\right]+D\Delta f_\pm+\frac{f_\pm-f_\mp}{\tau_s^e}=0.
\end{equation}
Here the two components corresponding to the subscript $\pm$ correspond to the electron spin parallel and antiparallel to the direction of $\bm\Omega_e$ [Eq.~\eqref{eq:Omega_e}], which is a good quantization axis at the time scale of the nuclear spin dynamics; $\bm\nabla=\partial/\partial\bm I$; $\Delta=\bm\nabla^2$; $\tau_s^{n,e}$ are the phenomenological nuclear and electron spin relaxation times unrelated with the hyperfine interaction, respectively; and $D=(\hbar\delta/A)^2/(2\tau_s^n)$ is the effective diffusion coefficient with $A$ being the hyperfine interaction constant. The nuclear spin dynamics mainly represents the precession with the frequency
\begin{equation}
  \label{eq:omega_n}
  \bm\omega_n^\pm=\pm\omega_e\frac{\bm\Omega_B}{\Omega_e}+\bm\omega_B,
\end{equation}
where $\omega_e=A/(2\hbar)$ is the nuclear spin precession frequency in the Knight field of completely spin polarized electron and $\bm\omega_B$ is the nuclear spin precession frequency in the external magnetic field.

The steady state solution of Eqs.~\eqref{eq:kinetic} has the form $f_\pm=f^{(0)}(\bm I)$, where
\begin{equation}
  f^{(0)}(\bm I)=\frac{1}{2}\left(\frac{A}{\sqrt{\pi}\hbar\delta}\right)^3\exp\left[-\left(\frac{AI}{\hbar\delta}\right)^2\right]
\end{equation}
in agreement with Eq.~\eqref{eq:F}.

Calculation of the angle $\theta_{12}$ in Eq.~\eqref{cur-form} for the given values of ${\bm I}^{(A)}$ and ${\bm I}^{(B)}$ shows that the current can be written as
\begin{equation}\label{bp-Fur}
  J_{AB} = {\cal J}_0 + {\cal J}_1\cos(\varphi) + {\cal J}_2\cos(2\varphi),
\end{equation}
where $\varphi$ is the polar angle between ${\bm I}^{(A)}$ and ${\bm I}^{(B)}$ and from Eqs.~\eqref{eq:Omega_e} and~\eqref{bp-Ts} we obtain
\begin{subequations}
  \label{bp-j012}
\begin{multline}\label{bp-j0}
{\cal J}_0 = J_0 + J_1\\
\times \frac{\left(O_b+I_z^{(A)}\right)^2\left(O_b+I_z^{(B)}\right)^2 + \left(I_{\perp}^{(A)} I_{\perp}^{(B)}\right)^2/2}{|{\bm O}_b + {\bm I}^{(A)}|^2 |  {\bm O}_b + {\bm I}^{(B)}|^2},
\end{multline}
\begin{equation}\label{bp-j1}
{\cal J}_1 = J_1
 \frac{2\left(O_b+I_z^{(A)}\right)\left(O_b+I_z^{(B)}\right) I_{\perp}^{(A)} I_{\perp}^{(B)} }{|{\bm O}_b + {\bm I}^{(A)}|^2 |   {\bm O}_b + {\bm I}^{(B)}|^2},
\end{equation}
\begin{equation}\label{bp-j2}
{\cal J}_2 = J_1
 \frac{\left(I_{\perp}^{(A)} I_{\perp}^{(B)}  \right)^2}{2|{\bm O}_b + {\bm I}^{(A)}|^2 |   {\bm O}_b + {\bm I}^{(B)}|^2}.
\end{equation}
\end{subequations}
Here $J_0$ and $J_1$ are given by Eqs.~\eqref{bp-JAB} or Eqs.~\eqref{eh-J-12} depending on the mechanism, $O_b = \hbar\Omega_B/A$ is a dimensionless frequency of the electron spin precession in the external magnetic field directed along $z$ axis, and $I_{\perp}^{(A,B)}$ are the absolute values of the spin components in the $(xy)$ plane.

The three contributions in Eq.~\eqref{bp-Fur} lead to the three independent contributions to the current noise spectrum. We assume that the typical nuclear spin precession frequency $\omega_e$ and electron spin relaxation rate $1/\tau_s^e$ are much larger than the nuclear spin relaxation rate $1/\tau_s^n$. In this case, the contribution related to ${\cal J}_0$ represents the low frequency noise at frequencies of the order of $1/\tau_s^n$. The two other terms give rise to the high frequency noise at the frequencies of the order of $\omega_e$. It is caused by the nuclear spin precession in the Knight field. Accordingly, the current noise spectrum can be written as
\begin{equation}\label{noise_dec}
(\delta J^2)_\omega = (\delta J^2)_\omega^{(HF)} + (\delta J^2)_\omega^{(LF)}.
\end{equation}
Below we separately calculate these two contributions.

\subsection{High frequency noise}
\label{HFbp}

The reason for the high frequency noise is the rotation of the nuclear spin in the Knight field of electron or hole localized at the given site. This rotation occurs only when the site is singly occupied. At unoccupied and doubly occupied sites there is not Knight field.

For the bipolaron mechanism, we assume that the number of the current-carrying electrons is much smaller than the number of hopping sites, so their effect on nuclear spins can be neglected. Nevertheless, one electron is always present at the site $B$ in the critical pair. It leads to the precession of the total nuclear spin ${\bm I}^{(B)}$ at the $B$ site around the magnetic field. In the same time, the nuclear spin ${\bm I}^{(A)}$ at the $A$ site is static neglecting the slow nuclear spin relaxation.

In the electron-hole mechanism  there are no resident change carriers at the sites $L$ and $H$. Nevertheless, the high frequency current noise takes place if one of the sites $L$ or $H$ acts as a trap for the charge carriers. It means that this site is almost always occupied. For the site $H$ the condition of being a trap is $W_{H}^{in} \gg \widetilde{\gamma}, \omega_n, W_H^{out}$. This condition ensures that the site $H$ is occupied almost immediately after losing its hole due to recombination or charge exchange with the contact.    In this case, it represents an analog of $B$ site. The other site should be analogous to $A$ site in the bipolaron mechanism, its occupation probability should be small.  Under these assumptions the high frequency noise in the bipolaron and electron-hole mechanism is the same. To be specific, we will use the notations of the bipolaron mechanism.

During the transmission of electron through the critical pair, the two electrons form a singlet spin state at $B$ site. So after the fast transmission, the electron at the $B$ site gets depolarized. The corresponding spin relaxation rate is
\begin{equation}
\frac{1}{\tau_s^{e}} = W_B^{in} + W_A^{in} \frac{W_{BA}}{W_{BA} + 2 W_{A}^{out}}.
\end{equation}
It defines the phenomenological spin relaxation time for the $B$ site. In the case of the electron-hole mechanism the spin relaxation time equals to $1/W_H^{out}$ or $1/W_L^{out}$ when $H$ or $L$ site represents a trap, respectively.

At the time scales much shorter than the nuclear spin relaxation time, the nuclear spin dynamics can be simply described by the two coupled Bloch equations:
\begin{equation}
  \label{eq:dIpm}
  \frac{\d\bm I^\pm}{\d t}=\bm\omega_n^\pm\times\bm I^\pm+\frac{\bm I^\mp-\bm I^\pm}{2\tau_s^e},
\end{equation}
where
\begin{equation}
  \bm I_\pm=\int f_\pm(t,\bm I)\bm I\d\bm I
\end{equation}
represents the average nuclear spins in the corresponding electron spin subspaces. Moreover, it is convenient to use the coordinate frame rotating with the frequency $\bm\omega_B$, because the electric current depends only on the angle between nuclear spins at the critical pair of sites. We assume that $\bm\omega_B$ at these sites is the same. As a result one can neglect the nuclear spin dynamics at the A site as well as the frequency $\omega_B$ at the B site.

Let us consider the current noise related to ${\cal J}_1$ in Eq.~\eqref{bp-Fur}. It is useful to introduce the correlation functions $c_{\sigma,\sigma'}=\langle e^{i\varphi(t)}\cos\varphi(0) \rangle_{\sigma,\sigma'}$, where $\sigma,\sigma=\pm$ correspond to the orientation of the electron spin at the $B$ site parallel or antiparallel to the $z$ axis at time moments $t$ and $0$, respectively. To calculate the average the expression in brackets should be multiplies by the corresponding spin projector operators at times $0$ and $t$. The correlators obey the equations (at $t>0$)
\begin{equation}\label{bp-cpm}
\frac{d}{dt}c_{\pm,\sigma} = \pm i\omega_n^{(B)}c_{\pm,\sigma} + \frac{c_{\mp,\sigma} - c_{\pm,\sigma}}{2\tau_s^{e}},
\end{equation}
according to Eq.~\eqref{eq:dIpm}. The initial conditions are $c_{\sigma,\sigma'}(0) = \delta_{\sigma,\sigma'}/4$. From the solution of these equations, the contribution to the current noise can be obtained as
\begin{equation}
  {\cal J}_1^2\sum_{\sigma,\sigma'}\int\limits_{-\infty}^\infty\Re\left[c_{\sigma,\sigma'}(t)\right]\e^{\i\omega t}\d t.
\end{equation}
In the same way we obtain the contribution ratio ${\cal J}_2^{(AB)}$. The total high frequency noise reads
\begin{multline}\label{noise-JAB}
(\delta J^2)_\omega^{(HF)} = \left\langle
{\cal J}_1^2 \frac{\tau_s^{e}\left(\omega_n^{(B)}\right)^2}{\omega^2 + \left(\tau_s^{e}\right)^2\left[\omega^2 - \left(\omega_n^{(B)}\right)^2\right]^2} \right.
\\
\left. + {\cal J}_2^2  \frac{4\tau_s^{e} \left(\omega_n^{(B)}\right)^2}{\omega^2 + \left(\tau_s^{e}\right)^2 \left[\omega^2 - 4\left(\omega_n^{(B)}\right)^2\right]^2}
\right\rangle,
\end{multline}
where the angular brckets denote only the averaging over the initial nuclear fields distribution. We perform it numerically. Note that the relation ${\cal J}_2/{\cal J}_1$ does not depend on hopping rates because both ${\cal J}_1$ and ${\cal J}_2$ are proportional to $J_1$, so the shape of the spectrum depends on the hopping rates through $\tau_s^{e}$ only.

\begin{figure}[t]
    \centering
        \includegraphics[width=\linewidth]{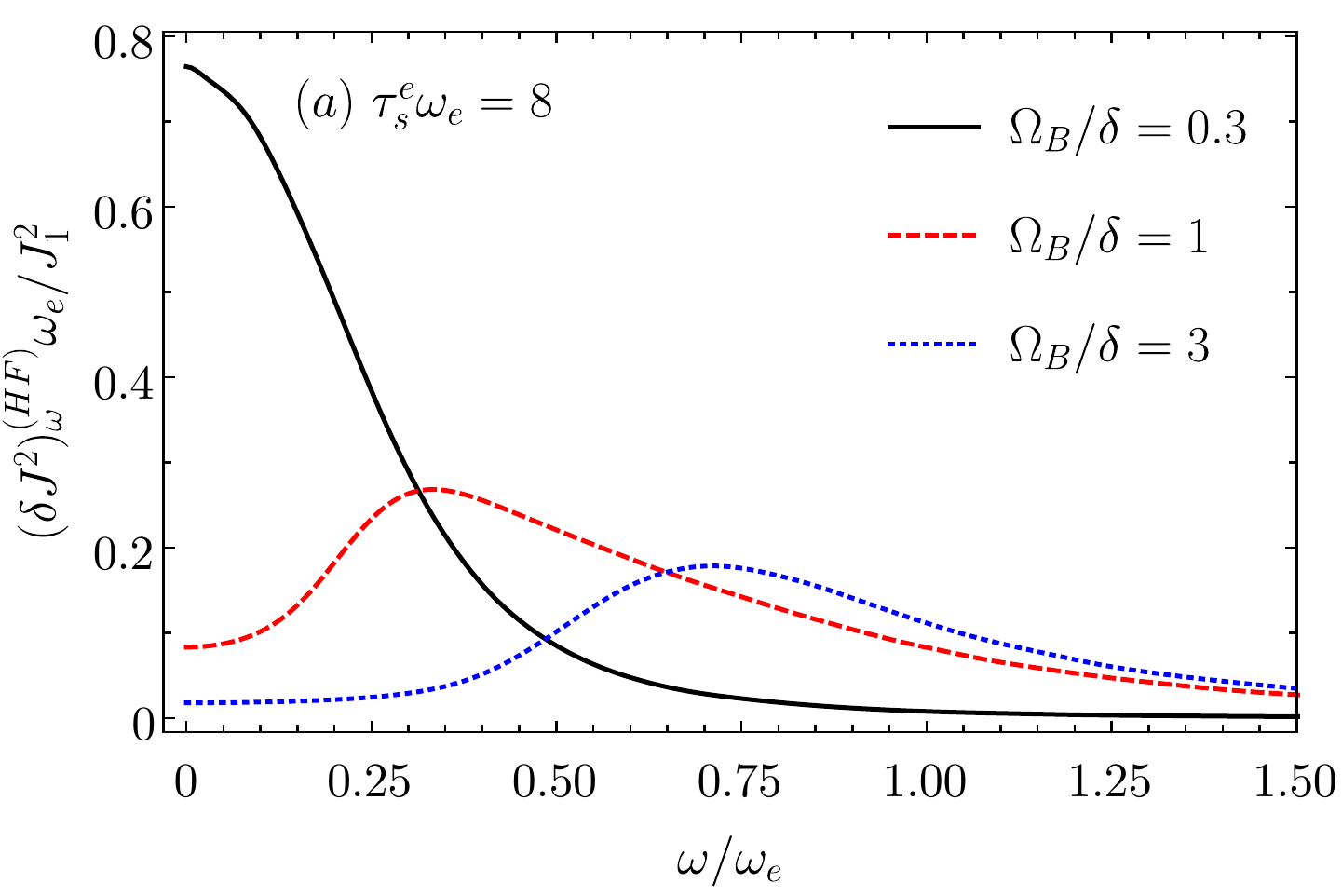}
        \includegraphics[width=\linewidth]{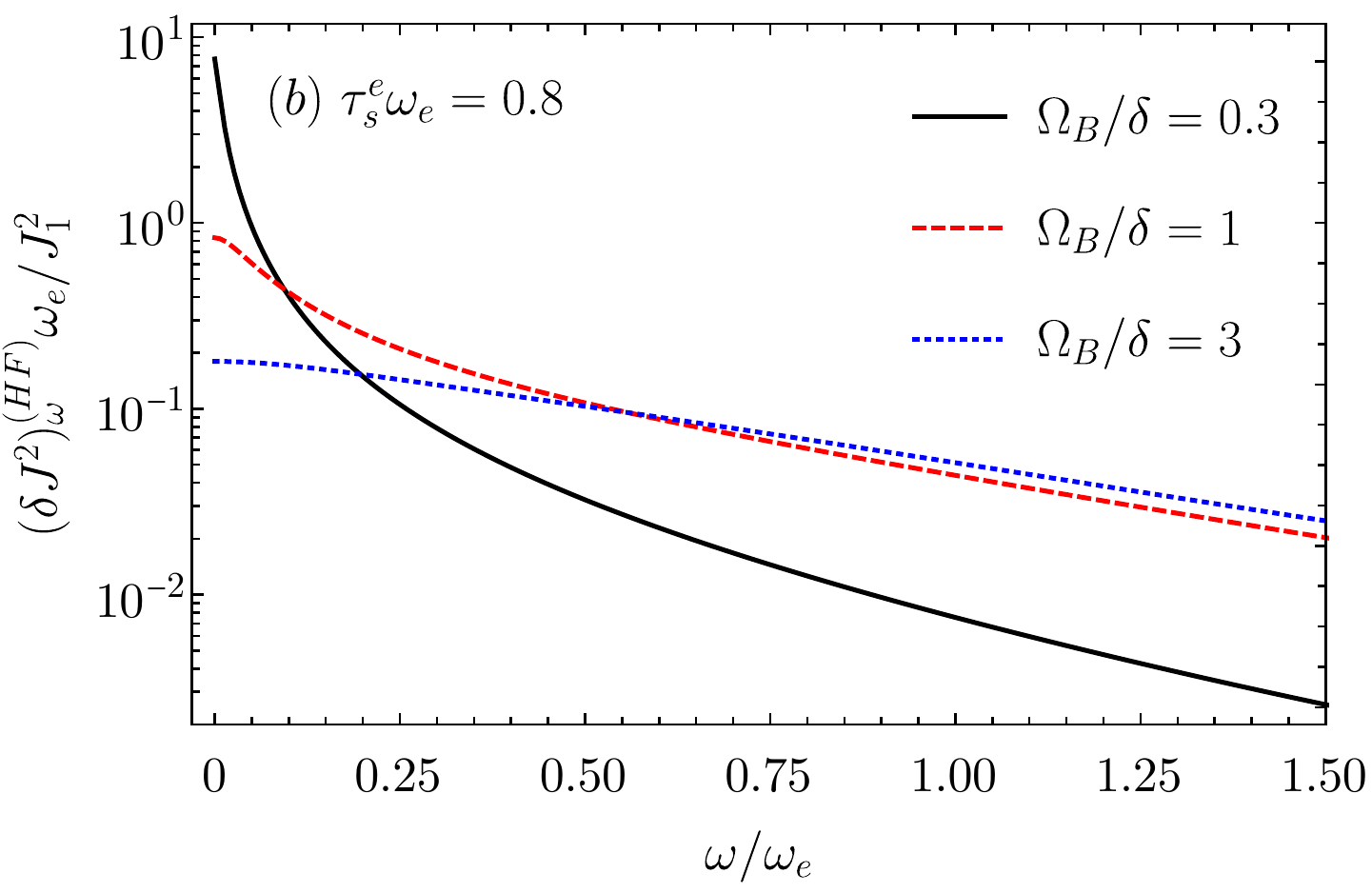}
        \caption{The high frequency current noise spectra calculated after Eq.~\eqref{noise-JAB} for the different strength of the magnetic field and different spin relaxation times, as indicated in the plots.}
    \label{fig:curnoise}
  \end{figure}

The high frequency contribution to the current noise spectrum in the bipolaron mechanism is shown in Fig.~\ref{fig:curnoise}. In the limit $\tau_s^{e}\gg1/\omega_e$, the shape of the spectrum does not depend on the hopping rates. This limit is illustrated in panel (a). The spectrum consists of a single asymmetric peak, which shifts to higher frequencies with increase of the magnetic field. Its central frequency saturates at $\omega = \omega_e$ in high fields. The additional peak at the frequency $2\omega_e$, which can be expected from Eq.~\eqref{noise-JAB} is very small and can not be clearly seen. The current noise intensity decreases with increase of the magnetic field. This is caused by the saturation of OMAR in large fields, when the current becomes independent of the orientation of the nuclear spins. The short electron spin relaxation time $\tau_s^{e}\lesssim1/\omega_e$ leads to the smearing of the noise spectrum, as shown in Fig.~\ref{fig:curnoise}(b).

\subsection{Low frequency noise}
\label{LFbpeh}

The low frequency noise is related to the nuclear spin components along the magnetic field, because they are conserved during the spin precession. Their dynamics is caused by the nuclear spin relaxation and leads to the current noise at the frequencies of the order of $1/\tau_s^n$.

The low frequency noise stems from the contribution ${\cal J}_0$ in Eq.~\eqref{bp-Fur}, which does not depend on the polar angles of nuclear spins, as it can be seen from Eq.~\eqref{bp-j0}. Therefore, this contribution can be described accounting for the diffusion related part of kinetic equation~\eqref{eq:kinetic} only. It is convenient to solve the diffusion equation using the fictitious Langevin forces (for each site):
\begin{equation}
  \frac{\d\bm I}{\d t}=\bm\xi(t)-\frac{\bm I}{\tau_s^n},
\end{equation}
with the correlation function
\begin{equation}
  \label{eq:xis}
 \braket{\xi_\alpha(t)\xi_\beta(t')}=\frac{1}{\tau_s^n}\left(\frac{\hbar\delta}{A}\right)^2\delta_{\alpha\beta}\delta(t-t'),
\end{equation}
$\delta_{\alpha\beta}$ is the Kronecker symbol and $\delta(t)$ is the Dirac delta function. As a result the shape of the low frequency current noise spectrum does not depend on the electron spin relaxation time and the hopping rates.

\begin{figure}[t]
  \centering
  \includegraphics[width=0.5\textwidth]{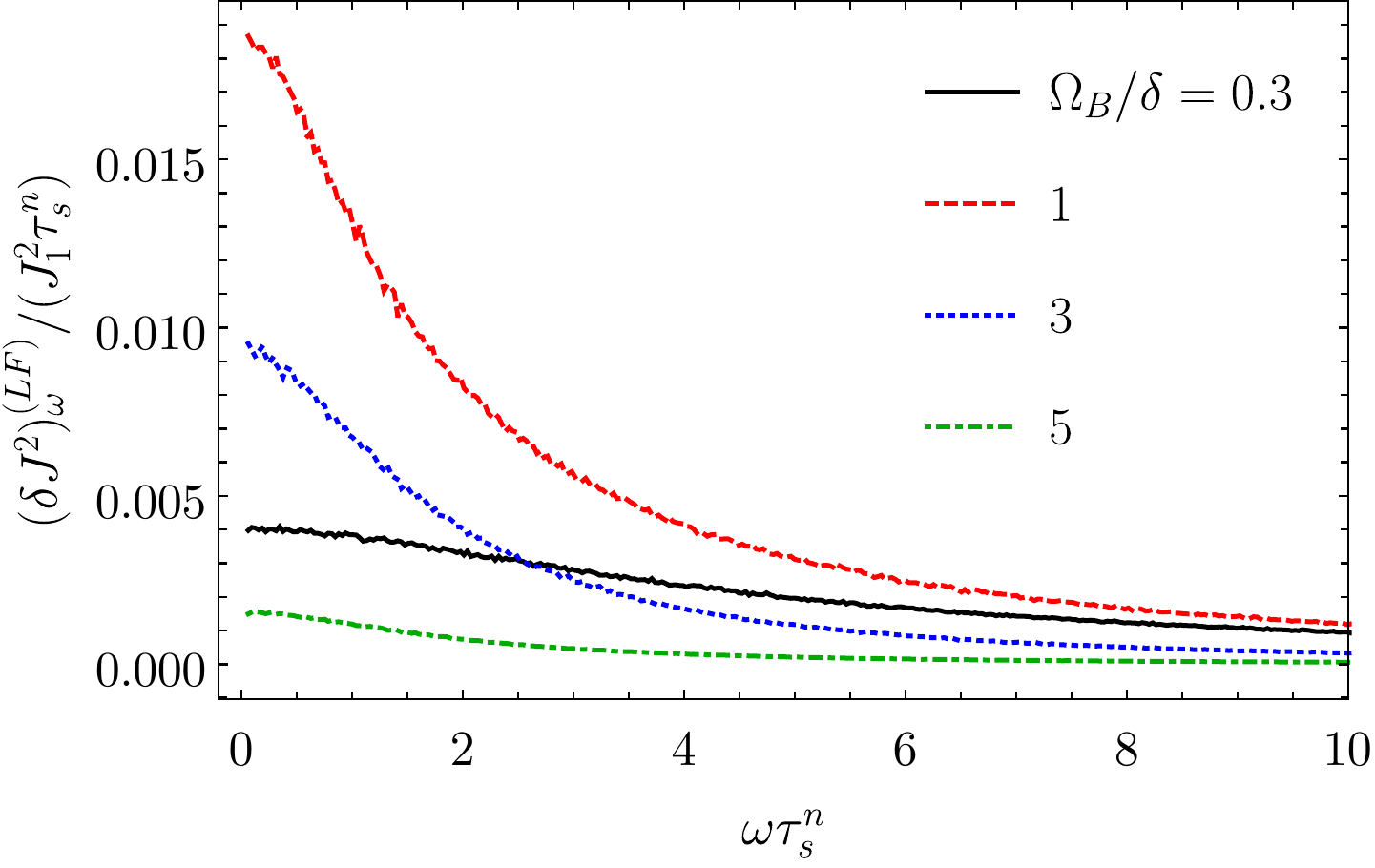}
  \caption{Low frequency current noise spectrum simulated numerically for the different magnetic fields.}
  \label{fig:zeta}
\end{figure}

We simulated low frequency noise numerically generating $10^3$ times the random forces for the time intervals $10^3\tau_s^n$. The results of the simulation are shown in Fig.~\ref{fig:zeta}. The spectrum always represents a peak at zero frequency with the width of the order of $1/\tau_s^n$. The amplitude of the peak nonmonotonously depends on the magnetic field. With increase of the field, it first increases and then decreases. Qualitatively, this happens because the low frequency noise can be viewed as a result of effective slow variations of the external magnetic field. From Fig.~\ref{fig:Dawson} one can see that these variations leads to the largest changes in the current when $\Omega_B\sim\delta$ in agreement with Fig.~\ref{fig:zeta}.

\section{Discussion and conclusion}
\label{sec:concl}

In our work we described the current noise in mesoscopic organic semiconductors, where it is the largest. When the size of the sample is larger than the correlation length, the number of the critical pairs in the percolation cluster can be estimated as $(L/L_c)^{d}$, where $d=2,3$ is the dimension of the sample~\cite{Shk}. We note that in this case the current noise is suppressed by the factor $(L_{c}/L)^{d/2}$.

In the experiments, the nuclei related current noise can be separated from the other contributions, such as shot noise and $1/f$ noise due to its sensitivity to the external magnetic field. We note, that the noise spectrum gets strongly modified in the same range of magnetic fields field, as OMAR does. In addition to this, the amplitude and the shape of the low-frequency peak described in Sec.~\ref{LFbpeh} can be controlled by the radiofrequency (RF) field. If it is applied in resonance with $\omega_B$, it increases the nuclear spin relaxation rate as~\cite{book_Glazov}
\begin{equation}
\frac{1}{\tau_s^n} = \frac{1}{\tau_s^{n(0)}} + \frac{\widetilde{\omega}^2}{2} \frac{\tau_s^{n(0)}}{1 + (\omega_{RF} - \omega_B)^2 (\tau_s^{n(0)})^2},
\end{equation}
where $\tau_s^{n(0)}$ is the spin relaxation time in the absence of the RF field, $\widetilde{\omega}$ is the nuclear spin precession frequency in the RF field, and $\omega_{RF}$ is the carrying frequency of the RF field. Thus application of the RF field leads to the broadening of the low frequency peak and decrease of its amplitude. The resonance dependence on $\omega_{RF}$ allows for unambiguous evidence of the role of nuclei in the OMAR.


In the case when OMAR is controlled by the electron-hole mechanism, the electron hole recombination is often radiative~\cite{kalin}. We note that in this case, the current noise can be conveniently measured by the fluctuations of the electroluminescence intensity.

In conclusion, we calculated the current noise in mesoscopic organic semiconductors caused by the nuclear spin fluctuations. This effect takes place in the samples with pronounced OMAR. The current noise spectrum consists of the two peaks. One is centered at the frequency, which increases with increase of the magnetic field and the other one is centered at zero frequency. The dependence of the nuclei induced current noise on the magnetic field as well as on the RF field allows one to separate its contribution to the current noise experimentally from the other contributions.

\section*{Acknowledgments}

We gratefully acknowledge the fruitful discussions with M. M. Glazov and the partial financial by the RF President Grant No. MK-1576.2019.2 and Foundation for the Advancement of Theoretical Physics and Mathematics ``Basis''. The calculation of the current noise spectrum by D.S.S. was supported by the Russian Science Foundation Grant No. 19-72-00081. A.V.S. acknowledges the support from the Russian Foundation for Basic Research Grant No. 19-02-00184.


%

\end{document}